\documentclass[12pt,showpacs,amsmath,amssymb]{revtex4}
\usepackage{graphicx} 
\usepackage{epsfig} 

\begin{document}

\title{Optimal behavior of viscoelastic flow at resonant frequencies}

\author{A.A. Lambert\footnote{ala@cie.unam.mx}, 
G. Ib\'{a}\~{n}ez\footnote{gid@cie.unam.mx}, S. Cuevas\footnote{scg@cie.unam.mx}
 and J.A. del R\'{\i}o\footnote{antonio@servidor.unam.mx}}

\affiliation{Centro de Investigaci\'{o}n en Energ\'{\i}a \\
Universidad Nacional Aut\'onoma de M\'exico\\
A.P. 34, 62580 Temixco, Mor. M\'exico }

\begin{abstract}

The global entropy generation rate in the zero-mean oscillatory
flow of a Maxwell fluid in a pipe is analyzed with the aim at
determining its behavior at resonant flow conditions. This
quantity is calculated explicitly using the analytic expression
for the velocity field and assuming isothermal conditions. 
The global entropy generation rate shows well-defined peaks
at the resonant frequencies where the flow displays maximum
velocities. It was found that resonant frequencies can be
considered optimal in the sense that they maximize the power
transmitted to the pulsating flow at the expense of maximum
dissipation.

\end{abstract}
\date{\today}

\maketitle

\section{INTRODUCTION}

There are several interesting phenomena where the existence of an
oscillatory flow leads to the improvement of a transport process
\cite{beamish}. For instance, the axial dispersion of contaminants
within laminar oscillatory flows in capillary tubes is considerably
larger than that obtained by pure molecular diffusion in the
absence of flow \cite{Chatwin1975,Watson1983}. Likewise, Kurzweg
\cite{Kurzweg-JHT,Kurzweg-JFM} found that in a zero-mean
oscillatory flow of a Newtonian fluid in a duct, the effective
thermal diffusivity reaches a maximum for a given oscillation
frequency. This leads to an enhanced longitudinal heat transfer
which involves no net convective mass transfer. In turn, it has
been found \cite{timp96, PRE58, PRE63} that the dynamic
permeability of a viscoelastic fluid flowing in a tube can be
substantially enhanced at specific resonant oscillation
frequencies. Under certain conditions, an enhanced flow rate
can be achieved. These phenomena may find important applications
in areas of technological interest such as nuclear reactors,
combustion processes and oil recovery \cite{patentoil,patentkur},
as well as for the understanding of physiological flows such as
those present in respiratory and circulatory systems
\cite{IJHMT40,SSEN00}. 

In this paper, we are interested in the relation between the 
irreversible behavior of an oscillating flow and the optimal
characteristics of the enhanced transport.
In recent years, a variety of systems have been analyzed and 
optimized using the Entropy Generation Minimization (EGM) method 
\cite{BEJAN94, BEJAN-EGM, CIHAT97, IJHMT46}. This method has 
become a useful tool for evaluating the intrinsic irreversibilities 
associated with a given process or device. By determining the 
conditions under which the entropy generation rate is minimized, the 
operating conditions can be optimized by reducing the dissipation to 
a minimum consistent with the physical constraints imposed on the 
system. 
In fluid flow systems, friction is
one of the main mechanisms responsible for entropy generation,
therefore, we must invest useful work to push the fluid through
the pipe against the irreversible viscous dissipation. In 
 this work, the entropy generation rate is used to evaluate 
the intrinsic irreversibilities associated with an oscillatory 
viscoelastic flow. Some interesting applications of thermodynamic 
optimization have been proposed by Bejan in the context of pulsating 
flows \cite{IJHMT40,SSEN00}. In particular, he has shown that in
the respiratory system, the minimization of the mechanical
power requirements by the thorax muscles during the inhaling
and exhaling cycle corresponds to the longest inhaling and 
exhaling strokes possible, while in ejaculation, the
maximization of the mechanical power transmitted to the
ejected seminal fluid explains the existence of an optimal
bursting time interval. It has to be pointed out that these
works consider only the viscous dissipative behavior of
fluids. However, most of biological fluids present a
viscoelastic nature and improved calculations must also
reflect their elastic behavior. In fact, del R\'{\i}o
{\it et al.} \cite{PRE58} speculated that the human heart
beats at the {\it optimum} pumping frequency to produce a
maximum flow through arteries and veins according to the
viscoelastic properties of the blood. Recently, this resonant
behavior was experimentally observed in a study of the dynamic
response of a Maxwellian fluid \cite{PRE68}, where the
enhancement at the frequencies predicted by the theory
was proved. In turn, Tsiklauri and Beresnev \cite{david,TPM53}
included the effect of longitudinally oscillating tube walls and
obtained the analogue enhanced behavior. All these results have
motivated to explore the consequences of the enhancement of the
dynamic response of an oscillating viscoelastic fluid under
different conditions \cite{PRE64,RMF49,JASA112}.
At this point, the question whether this {\it optimum} pumping 
behavior is also optimum or efficient from a thermodynamical 
point of view can be formulated. This problem can be
addressed through the analysis of the entropy generation
rate \cite{BEJAN94, BEJAN-EGM}. In this paper, we have focused 
our attention in the analysis of the relationship between maximum
permeability (and, therefore, maximum velocity) of a zero-mean
oscillatory flow of a viscoelastic fluid in a rigid cylindrical
tube and the entropy generation rate that characterizes the
process. From the analytic expression for the velocity field,
shear stresses are determined and the local and global entropy
generation rate as a function of the oscillation frequency are
calculated. 
In this work, irreversibilities due to heat flow
phenomena are not considered.

\section{THEORETICAL MODEL}

We consider the flow of a Maxwell fluid in a rigid cylindrical 
tube of radius $a$ under an oscillatory pressure gradient 
applied in the longitudinal $x$-direction. This problem was 
solved analytically by del R\'{\i}o {\it et al.} \cite{PRE58} 
in the linear regime, and the corresponding velocity field 
$V(r,t)$ reads
\begin{equation}
   V\left(r,t\right) = -\frac{1+i \omega t_m}{\beta^2 \eta}\left[1 -
   \frac{J_0\left(\beta r\right)}{J_0\left(\beta a\right)}\right]
   \frac{dP}{dx},
   \label{1}
\end{equation}
where the no-slip condition has been imposed at the wall of the 
cylinder, $V(a)=0$. Here 
$\beta=\sqrt{(\rho/\eta t_m)\left[\left(t_m \omega \right)^2 -i 
\omega t_m \right]}$, $\eta$ and $\rho$ are the dynamic viscosity 
and mass density of the fluid, $t_{m}$ is the relaxation time for 
the Maxwell fluid, $J_0$ is the cylindrical Bessel function of 
zeroth order and $dP/dx$ is the general expression of the 
time-dependent pressure gradient. All physical properties of 
the fluid are considered constant. In order to obtain analytical 
results, in this work we chose a harmonic pressure gradient given
by the real part of the expression $P_x e^{-i \omega t}$, where 
$P_x$ is the constant amplitude of the pressure gradient and 
$\omega$ is the angular frequency. With this assumption, the 
dimensionless expression for the velocity field is
\begin{equation}
V^{\ast}(r^{\ast},t^{\ast})=-\frac{1+i\omega ^{\ast}}{\alpha \varpi}
\left( 1-
\frac{J_{o}\left( \sqrt{\alpha \varpi }r^{\ast }\right) }{J_{o}\left(
\sqrt{\alpha \varpi }\right) }\right) e^{-it^{\ast }},
\label{2}
\end{equation}
where $V^{\ast}$, $\omega ^{\ast}$,  $r^{\ast}$ and $t^{\ast}$ 
have been normalized by 
$V_{o}=(a^{2}/\eta )P_x$, $1/t_m$, $a$ and $1/\omega $, 
respectively. Here,
$\varpi =\left( \omega t_m\right) ^{2}-i\omega t_m$ while 
$\alpha=a^{2}\rho /\eta t_m$ is the Deborah number.
 
\subsection{Entropy Generation Rate}

We now proceed to calculate the entropy generation rate. 
Since the fluid is assumed to be a simple substance, mass 
diffusion phenomena are disregarded. In addition, we consider
that the main source of entropy generation is given by 
frictional effects. However, it is assumed that the rise in 
temperature in the fluid and walls due to this dissipative effect
is negligible so that temperature remains approximately constant 
and irreversibilities due to heat transfer are not taken into 
account. Under these approximations, the dimensionless
local entropy generation rate, $\dot{S}^{\ast }$, 
that characterizes the irreversible behavior of the system, is given by 
\begin{equation}
  \dot{S}^{\ast }(r^{\ast}, t^{\ast})=\frac{1}{T^{\ast }}\left( \frac{\partial V^{\ast }}
{\partial r^{\ast }}\right) ^{2},
\label{3}
\end{equation}
where $\dot{S}^{\ast }$ and the dimensionless temperature of 
the fluid, $T^{\ast}$, are normalized by $V_{o}^{2}\eta /T_{o}a^{2}$ 
and $T_o$, respectively, $T_{o}$ being the mean dimensional fluid 
temperature. Notice that $V^{\ast}$ and $\dot{S}^{\ast}$ always 
are in phase, the temporal variation being $cos (t^{\ast})$ 
and $cos^{2} (t^{\ast})$, respectively. 
In order to obtain the entropy generation rate per unit length in the
axial direction, $<S^{\ast }>$, $S^{\ast }$ is integrated over the tube
cross-section.
Thus, $<\dot{S}^{\ast}>$ is only a function of $t^{\ast}$,  
$\omega ^{\ast}$ and $\alpha$. 
The corresponding averaged velocity over the tube cross-section is
\begin{equation}
<V^{\ast}>= \frac{2\pi}{A} \int^{1}_{0} V^{\ast}
(r^{\ast},t^{\ast})r^{\ast} dr^{\ast}.
\label{21}
\end{equation}
where $A$ is the cross-section area. 
We can now use equations (\ref{3}) and (\ref{21}) to
characterize the resonant behavior of the system.

\section{RESULTS}

\begin{figure}
\begin{center}
\includegraphics[width=0.8\hsize,angle=0]{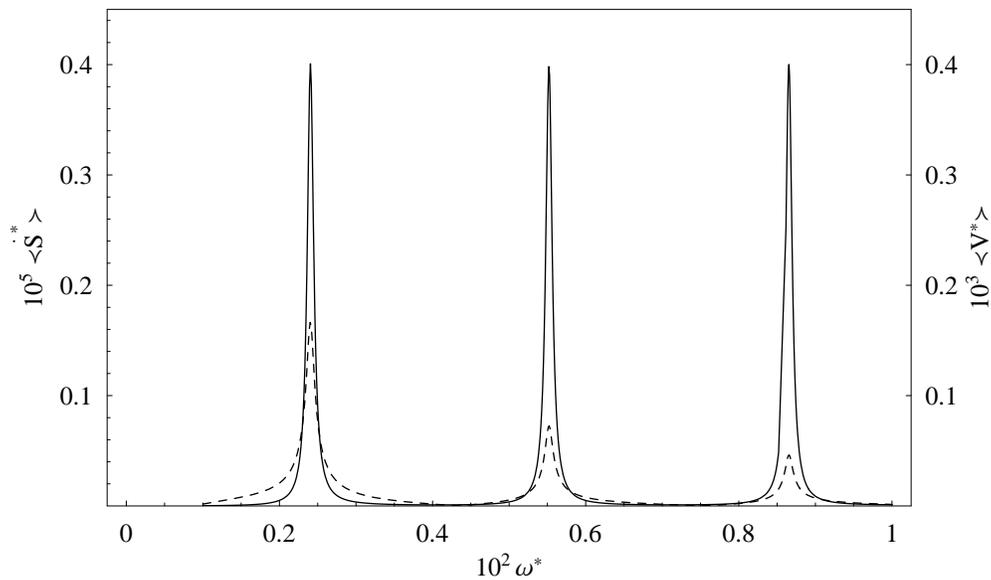}
\caption{The amplitudes of the dimensionless velocity $<V^{\ast}>$ 
(dashed line) and global entropy generation rate 
$<\dot{S}^{\ast}>$ (solid line) as a function of the
dimensionless frequency $\omega ^{\ast}$ with $\alpha=0.01$.}
\label{fig1}
\end{center} 
\end{figure}

In Fig. \ref{fig1}, the amplitudes of the averaged velocity and the global
entropy generation rate are shown as a function of the
dimensionless frequency  for a Deborah number $\alpha=0.01$.
For comparison purposes, we have used the same value of 
$\alpha$ as in the paper by del R\'{\i}o, {\it et al.}
\cite{PRE58}. It corresponds to a fluid with a relaxation
time of the order of seconds, a mass density and viscosity
of the same order of water, and a tube radius of the order
of centimeters. With this value, viscoelastic behavior is
well established. In fact, physical properties of
cetylpyridinium chloride and sodium salicylate solution
(CPyCl/NaSal, $60$/$100$)~\cite{Micelas,Hoffman} give an
$\alpha$ value close to $0.01$. For simplicity, in all 
calculations presented here, the dimensionless temperature
was taken as $T^{\ast}=1$. Notice that the maximum values of
$<\dot{S}^{\ast}>$ are found at the resonant 
frequencies where $<V^{\ast}>$ is also maximum.
This has important implications in terms of the useful work
that is invested to move the fluid through the pipe: maximum 
velocity is obtained at the expense of maximum dissipation.
On the other hand, from the relationship between work $W$
and velocity $v$, namely, $dW/dt=P A v$, it is clear that 
for a given pressure $P$ and cross-sectional area $A$,
maximum fluid velocity leads also to maximum power.
Therefore, it follows that resonant frequencies can be 
considered optimal in the sense that they maximize 
the power transmitted to the fluid through the pulsating 
flow. 
 
An interpretation of this result in terms of Darcy's law can also 
be given. The phenomenological law for a frequency-dependent mean 
flux (or average velocity) can be expressed as 
${\bf J}=<\textbf{V}^{\ast}>= -K(\omega^{\ast}) \nabla P$, where 
$K(\omega^{\ast})$ is the dynamic permeability \cite{PRE58}. 
Therefore, expressing the global entropy generation rate as 
the product of fluxes and generalized forces \cite{GROOT1984}, 
we get 
\begin{equation}
<\dot{S}^{\ast}>= -\frac{1}{T}{\bf J} \cdot \nabla P= 
\frac{K(\omega^*)}{T}  \mid \nabla P \mid^2
\label{4}
\end{equation}
where in order to satisfy the condition 
$<\dot{S}^{\ast}> \;\; \ge 0$, the dynamic permeability 
must be a positive definite quantity. From Eq. (\ref{4}), 
it is then clear that maximum values of $<\dot{S}^{\ast}>$ 
will be obtained at those frequencies at which the
$K(\omega^{\ast})$ is maximized. But from Darcy's law 
these are precisely the frequencies that lead to maximum 
mean flux or average velocity.

\begin{figure}
\begin{center}
\includegraphics[width=0.75\hsize,angle=0]{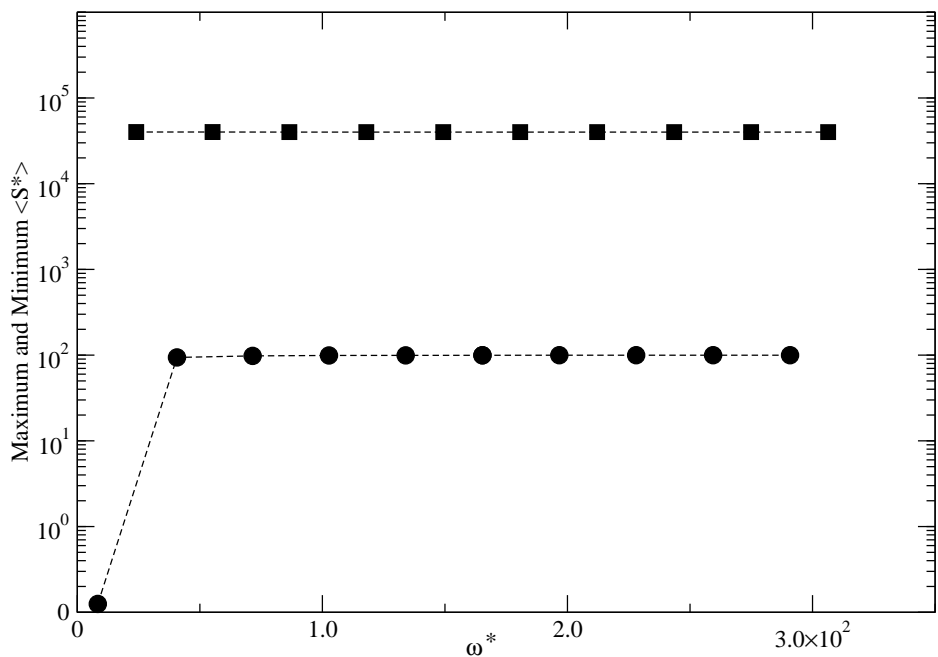}
\caption{Maximum (square) and minimum (dot) 
values of $<\dot{S}^{\ast}>$ at different resonant 
frequencies. The dashed lines show only the trend behavior.}
\label{fig2}
\end{center} 
\end{figure} 

It is also interesting to observe in Fig. \ref{fig1} that 
while maximum values of $<V^{\ast}>$ decrease as 
higher resonant frequencies are reached, maximum values 
of $<\dot{S}^{\ast}>$ remain almost constant. This 
is more clearly shown in Fig. \ref{fig2} where maxima 
and minima of the global entropy generation rate are 
presented as a function of the frequency. This result 
indicates the importance of the first resonant frequency where 
the higher mean velocity is obtained. The irreversibilities 
associated to the production of the first peak velocity are 
approximately the same as those involved in the production of 
the remaining peaks although maximum velocity values decrease 
the higher the frequency. A drastic rise in the minima is 
observed from zero frequency to the first minimum, but from 
that value the remaining local minima stay almost constant 
and, in fact, they reach a limit value as 
$\omega^{\ast} \rightarrow \infty$. It is important to 
emphasize the fact that the lowest minimum of the global 
entropy generation rate corresponds to a stationary state, 
{\it i.e.}, to the zero frequency. This result is in 
agreement with Prigonine's theorem, which states a 
minimum entropy generation for stationary states provided 
that the Onsanger coefficients are constant \cite{GROOT1984}.

\section{CONCLUSIONS}

In this paper, we have used the global entropy generation 
rate to analyze a zero-mean oscillatory flow of a Maxwell 
fluid at resonant conditions. It was found that the global 
entropy generation rate is maximized at the same frequencies 
at which the flow displays a resonant behavior. Under these 
conditions the average velocities are maximum and the 
power transmitted to the fluid through the pulsating flow 
is also maximum. Therefore, it is from the maximization of 
power that pumping at resonant frequencies can be considered 
optimal. However, from a thermodynamic point of view, maximum 
average velocities are reached through the maximization of flow
irreversibilities. Given the viscoelastic nature of most 
biological fluids, this may help to the understanding of some 
pulsating physiological processes \cite{IJHMT40,SSEN00,PRE58}. 

It was observed that global entropy generation rate remains 
the same at different resonant frequencies although the 
maximum values of velocity decrease at higher frequencies.
On the other hand, the existence of a lowest minimum value 
of $<\dot{S}^{\ast}>$ is in agreement with Prigonine's 
theorem of minimum entropy production for stationary states.

\strut

{\bf ACKNOWLEDGMENTS}

This work was supported partially by CONACYT under Project No. 38538. A.A. 
Lambert thankfully acknowledges financial support from UABC and PROMEP-SEP.

\end{document}